# Relative merits of Phononics vs. Plasmonics: the energy balance approach.

Jacob B Khurgin

Johns Hopkins University

Baltimore MD 21218 USA

**Abstract**

The common feature of various plasmonic schemes is their ability to confine optical fields of surface plasmon polaritons (SPPs) into sub-wavelength volumes and thus achieve a large enhancement of linear and nonlinear optical properties. This ability, however, is severely limited by the large ohmic loss inherent to even the best of metals. However, in the mid and far infrared ranges of the spectrum there exists a viable alternative to metals – polar dielectrics and semiconductors in which dielectric permittivity (the real part) turns negative in the *Reststrahlen region*. This feature engenders the so-called surface phonon polaritons (SPhPs) capable of confining the field in a way akin to their plasmonic analogues, the SPPs. Since the damping rate of polar phonons is substantially less than that of free electrons, it is not unreasonable to expect that "*phononic*" devices may outperform their *plasmonic* counterparts. Yet a more rigorous analysis of the comparative merits of phononics and plasmonics reveals a more nuanced answer, namely that while phononic schemes do exhibit narrower resonances and can achieve a very high degree of *energy* concentration, most of the energy is contained in the form of lattice vibrations so that enhancement of the *electric field,* and hence the Purcell factor, is rather small compared to what can be achieved with metal nanoantennas. Still, the sheer narrowness of phononic resonances is expected to make phononics viable in applications where frequency selectivity is important.

1. Introduction

The discipline of optics for most of its existence has relied on a very limited variety of optical materials, mostly dielectrics with a narrow range of refractive indices, from about 1.38 for $MgF_2$ to about 2.4 for $TiO_2$ in the visible and 4 for Ge in the near and mid-infrared [1]. Consequently, the minimum size of optical components has been historically limited by the diffraction limit to about $\lambda/n$, i.e. a few hundred nanometers, making the density of optical integration much lower than the density of electronic integration. In the course of the last decade and a half the situation slowly started to change as the palette of optical materials has been expanded to include materials with negative permittivity (i.e. imaginary refractive index), including mostly metals but also doped semiconductors. As a result a new discipline, plasmonics, has emerged, free of the constraints imposed by the diffraction limit [2-3]. At about the same time, the optics community realized that by combining sub-wavelength parts made from materials with different signs of permittivity, entirely new artificial media with properties unattainable in natural materials can be synthesized.

These media have been named "metamaterials" and a number of exciting potential applications for them, ranging from superlensing to optical cloaking, have emerged [4-6]. However, after a few giddy years of unlimited promise the research in metamaterials and plasmonics has hit a wall as the community has started to recognize the obvious fact that ohmic loss in the metal prevents plasmonic and metamaterial devices from properly performing their functions [7,8]. Once this unfortunate yet unescapable fact settled in, the research has slowly migrated towards areas where the loss may not be the deciding factor (plasmonic sensors) [9] or where it can in fact be useful (photo-catalysis, thermal photovoltaics and others [10-13]). It was also suggested that loss can be mitigated by optical amplification [14-17], but soon it was shown that the loss is simply too large for that and introduction of a gain medium only increases noise [18-20]. Hence, the glimmer of hope for plasmonics and metamaterials use in such applications as integrated optics still remains in discovering or developing new low loss material systems.

While in the visible and near IR range it is very difficult to find an alternative to the noble metal (and not for the lack of trying) the situation in the mid-IR –to-THZ range appears to be far more promising [21-26]. The scattering rates in highly doped semiconductors are on the order of $10^{13} s^{-1}$, i.e. about an order of magnitude less than in the metals, however the plasma frequency is also significantly less, i.e. the skin depth in semiconductors is much larger than in metals. As the field penetrates deep inside the semiconductors it gets absorbed and the resulting effective loss in the doped semiconductors is actually larger than in metals [27-29].

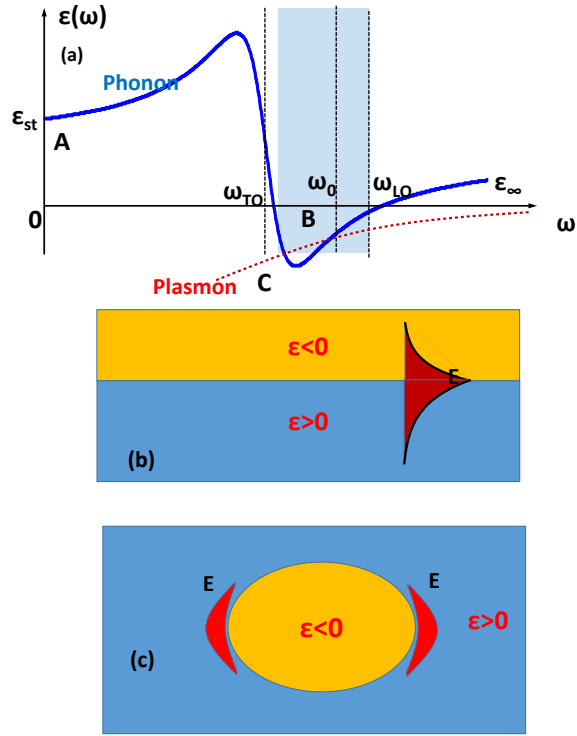

Figure 1 (a) Dispersion of phonons and plasmons. The regions A, B, and C correspond to three different ways to achieve a resonant mode: dielectric resonator, SphP, and SPP respectively. (b) Propagating SPhP; (c) Localized SPhP.

Nevertheless, there exists yet another pathway leading to extensively low loss, negative permittivity materials in the IR region – using the motion of ions in place of free carriers [30-39]. Just like the oscillations of free carriers in metal or semiconductors, the collective oscillations of ions in polar dielectrics (commonly referred to as optical phonons) engender oscillating space charges. These oscillations in turn can couple (hybridize) with the electro-magnetic field (photons) and the combined modes are known as phonon polaritons [40,41]. The optical phonons are characterized by the transverse optical phonon resonance frequency $\omega_{TO}(k)$ (typically, dependence on the wave vector $k$ is rather weak) and the dispersion of the dielectric constant can be modeled by the Lorentzian oscillator,

$$\varepsilon_r(\omega) = \varepsilon_\infty \left(1 + \frac{\omega_P^2}{\omega_{TO}^2 - \omega^2 - j\omega\gamma}\right) \quad (1)$$

where $\varepsilon_\infty$ is the high frequency (optical) dielectric constant of the material, $\gamma$ is the damping (or momentum relaxation) rate, and the plasma frequency $\omega_P^2 = Ne^{*2}/\varepsilon_0\varepsilon_\infty M_r$, where $N$ is the density of polar bonds, $e^*$ is their effective charge and $M_r$ is the reduced mass. The real part of the dielectric constant becomes equal to 0 at the frequency of $\omega_{LO} = \sqrt{\omega_{TO}^2 + \omega_P^2}$ called the frequency of longitudinal optical phonons. For a typical phononic material, SiC [42,43] $\omega_{TO} \approx 2\pi \times 23.9 THz$, $\omega_{LO} \approx 2\pi \times 29.2 THz$, $\omega_P \approx 2\pi \times 16.7 THz$ and $\gamma \approx 1.1 ps^{-1}$. Now, as one can see from Fig.1a, in the frequency range $\omega_{TO} < \omega < \omega_{LO}$ the real part of $\varepsilon_r$ is negative and the propagating electro-magnetic waves are not supported in this so-called *Reststrahlen* region. However the interface modes, called surface phonon-polaritons (SPhPs) do exist at the interface between the two dielectrics, one with positive and one with negative dielectric constants as shown in Fig.1b where the propagating SPhP is shown and Fig.1c depicting the localized SPhP mode. Both localized and propagating modes look very similar to their counterparts in the metal structures-surface plasmon polaritons (SPPs) which is not surprising given that the dispersion of the metal (or doped semiconductor) looks similar,

$$\varepsilon_{met}(\omega) = \varepsilon_\infty \left(1 - \frac{\omega_p^2}{\omega^2 + j\omega\gamma_f}\right) \quad (2)$$

and can be obtained from (1) by simply setting the TO resonance frequency to zero. The dispersion of the metal (or doped semiconductor) is also shown (not necessarily to scale) as a dotted line in Fig.1a

The most attractive feature of SPhPs, which has given impetus to the whole new field of "phononics", is that the damping rate $\gamma$ of optical phonons, caused by the anharmonicity, is typically as slow as $10^{12} s^{-1}$ (scattering time is on the scale of a picosecond). This compares very favorably with the damping rate of SPP's, caused by the phonon and surface assisted absorption $\gamma_f$, which happens to be on the order of $10^{14} s^{-1}$ in metals and $10^{13} s^{-1}$ in highly doped semiconductors. Hence it seems logical that in the IR region the SPhPs should possess multiple advantages over the SPP's in terms of the propagation length, lifetime, and the field enhancement. Indeed, a significant amount of work in phononics has been performed in the course of the last decade and many of the plasmonic experiments, previously performed in the visible range, have been successfully implemented in the Mid-IR region using phononic structures [31-39,44].

While the results obtained in the aforementioned references clearly demonstrate that narrow resonances are indeed achievable in phononic structures, it is not clear that the degree of field enhancement is superior to what can be attained with metals, or doped semiconductors. The presence of the resonance frequency $\omega_{TO}$ in the denominator of (1) vs. the absence of it in (2) is expected to make the difference, but how big this difference is and what are the physical reasons for it have not been investigated in detail. In this work we present a simple and physically transparent theory that elucidates the difference between phononic and plasmonic materials using nothing but the energy conservation considerations.

## 2. Energy balance in polar dielectric structures

Let us consider the energy balance inside a mode contained within some volume of dielectric with relative permittivity $\varepsilon_r(\omega)$. If the characteristic dimension of the volume is $a$, then the electric field can be written as roughly $E \sin(\pi x/a)\sin(\omega t)$ and magnetic field as $H \cos(\pi x/a)\cos(\omega t)$. Then from the Maxwell's equation $\nabla \times H = i\omega\varepsilon_0\varepsilon_r E$ one can obtain the order-or-magnitude relation between the magnitudes of two fields,

$$H \approx \frac{\omega a}{\pi}\varepsilon_0\varepsilon_r E = \frac{2a}{\lambda}\frac{\varepsilon_r E}{\eta_0} \quad . \tag{3}$$

The time-averaged electric energy density can be written as

$$\langle U_E \rangle = \frac{1}{4}\varepsilon_0 \frac{\partial(\omega\varepsilon_r')}{\partial\omega} E^2 \tag{4}$$

where $\varepsilon_r'$ is the real part of dielectric constant while the time-averaged magnetic energy density is

$$\langle U_M \rangle = \frac{1}{4}\mu_0 |H|^2 \sim \left(\frac{2na}{\lambda}\right)^2 \frac{1}{4}\varepsilon_0\varepsilon_r' E^2 \tag{5}$$

where $n = \text{Re}(\sqrt{\varepsilon_r})$. If one considers the lowest order mode in the cavity, with $a = \lambda/2n$, and neglects the dispersion one immediately obtains the energy conservation relation $\int\langle U_E\rangle d^3r = \int\langle U_M\rangle d^3r$.

But the time averaged picture does not properly represent the energy balance in the mode since the electric energy includes contributions oscillating 90 degrees out of phase with each other (in-phase and quadrature components). According to the Lorentz oscillator model, used to derive (1) the relative displacement of ions is

$$x(t) = \frac{e^*/M_r}{\left[\left(\omega_{TO}^2 - \omega^2\right)^2 + \omega^2\gamma^2\right]^{1/2}} E \sin(\omega t - \varphi) \qquad (6)$$

where $\tan\varphi = \gamma^2\omega^2 / (\omega_{TO}^2 - \omega^2)^2$, and the velocity of this motion is

$$\dot{x}(t) = -\omega \frac{e^*/M_r}{\left[\left(\omega_{TO}^2 - \omega^2\right)^2 + \omega^2\gamma^2\right]^{1/2}} E \cos(\omega t - \varphi) \ . \qquad (7)$$

Then we obtain the expressions for the kinetic

$$U_K(t) = \frac{1}{2} NM_r \dot{x}^2(t) = \frac{1}{2} \frac{\omega^2 \omega_P^2}{\left(\omega_{TO}^2 - \omega^2\right)^2 + \omega^2\gamma^2} \varepsilon_0 \varepsilon_\infty E^2 \cos^2(\omega t - \varphi) \qquad (8)$$

and potential

$$U_P(t) = \frac{1}{2} NM_r \omega_{TO}^2 x^2(t) = \frac{1}{2} \frac{\omega_{TO}^2 \omega_P^2}{\left(\omega_{TO}^2 - \omega^2\right)^2 + \omega^2\gamma^2} \varepsilon_0 \varepsilon_\infty E^2 \sin^2(\omega t - \varphi) \qquad (9)$$

energy densities inside the volume. The potential energy then can be split into two parts as

$$U_P(t) = \frac{1}{2} \frac{\omega_P^2(\omega_{TO}^2 - \omega^2 + \omega^2)}{\left(\omega_{TO}^2 - \omega^2\right)^2 + \omega^2\gamma^2} \varepsilon_0 \varepsilon_\infty E^2 \sin^2(\omega t - \varphi) = U_{P1}(t) + U_{P2}(t) \ . \qquad (10)$$

The first part is

$$U_{P1}(t) = \frac{1}{2} \frac{\omega_P^2(\omega_{TO}^2 - \omega^2)}{\left(\omega_{TO}^2 - \omega^2\right)^2 + \omega^2\gamma^2} \varepsilon_0 \varepsilon_\infty E^2 \sin^2(\omega t - \varphi) = \frac{1}{2} \varepsilon_0 \left(\varepsilon_r^{'} - \varepsilon_\infty\right) E^2 \sin^2(\omega t - \varphi) \qquad (11)$$

where $\varepsilon_r^{'}$ is the real part of the dielectric constant (1) and the second part

$$U_{P2}(t) = \frac{1}{2} \frac{\omega^2 \omega_P^2}{\left(\omega_{TO}^2 - \omega^2\right)^2 + \omega^2\gamma^2} \varepsilon_0 \varepsilon_\infty E^2 \sin^2(\omega t - \varphi) = \frac{1}{4} \varepsilon_0 \omega \frac{\partial \varepsilon_r^{'}}{\partial \omega} E^2 \sin^2(\omega t - \varphi) \qquad (12)$$

has the same amplitude as kinetic energy (8) but its phase is shifted by 90 degrees. The total electric energy density can then be found as

$$U_E(t) = U_{P1}(t) + U_{P2}(t) + U_\infty(t) + U_K(t) = \frac{1}{2} \varepsilon_0 \varepsilon_r^{'} E^2 \sin^2(\omega t - \varphi) + \frac{1}{4} \varepsilon_0 \omega \frac{\partial \varepsilon_r^{'}}{\partial \omega} E^2 \qquad (13)$$

where

$$U_\infty(t) = \frac{1}{2}\varepsilon_0\varepsilon_\infty E^2 \sin^2(\omega t) \approx \frac{1}{2}\varepsilon_0\varepsilon_\infty E^2 \sin^2(\omega t - \varphi) \qquad (14)$$

is the sum of energy stored in the electric field proper, $U_{EF} = \frac{1}{2}\varepsilon_0 E^2$, and the potential energy associated with oscillations of valence electrons, $U_V = \frac{1}{2}\varepsilon_0(\varepsilon_\infty - 1)E^2$. Neglecting a small phase shift $\varphi$ in (14) amounts to a very small error on the scale of $\gamma^2/\omega_p^2$. Averaging (13) over time immediately leads to (4).

Now we can see that the energy oscillating roughly in phase (i.e. as $\sin^2(\omega t - \varphi)$) with the electric field, which can be referred to as either "in phase" or "potential", is

$$U_I(t) = U_\infty(t) + U_{P1}(t) + U_{P2}(t) \qquad (15)$$

and has three components. The first of them, $U_\infty(t)$, is entirely static as it has no frequency dependence. The second component, $U_{P1}(t)$ is only weakly resonant and dominates the frequency response in the normal dispersion region. The third component, $U_{P2}(t)$, whose amplitude is equal to the amplitude of kinetic energy $U_K(t)$, is very dispersive and becomes the dominant factor in the anomalous dispersion region. This breakdown is shown in Fig.2 a and c.

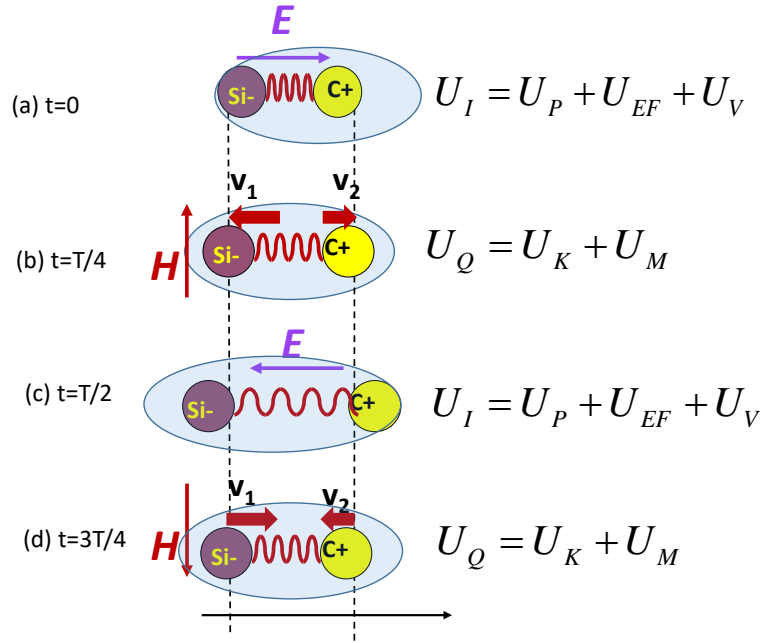

Figure 2 Energy breakdown in the polar dielectric material excited by the electro-magnetic field for four different times within oscillation period T.

Note that in the anomalous region when $\omega > \omega_{TO}$, $U_{P1}(t)$ becomes negative but the total potential energy (10) is of course always positive. The energy that oscillates roughly 90 degrees out of phase with the electric field (i.e. as $\cos^2(\omega t - \varphi)$) can be referred to as either "quadrature" or kinetic in the Lagrangian mechanics sense

$$U_Q(t) = U_K(t) + U_M(t) \tag{16}$$

and it has two components, the actual kinetic energy of the ions, $U_K(t)$ which is strongly dispersive, and the magnetic energy, $U_M(t)$, as shown in Fig.2 b and d. Whether kinetic of magnetic energy dominates depends on the dimensions of the mode. Comparing (8) with (5) immediately yields

$$\frac{U_K}{U_M} \sim \left(\frac{\lambda}{2a}\right)^2 \frac{\omega}{2|\varepsilon_r'|^2} \frac{\partial \varepsilon_r'}{\partial \omega} \sim \frac{1}{2}\left(\frac{\lambda}{2a}\right)^2 \left(\frac{\omega^2}{\varepsilon_\infty \omega_P^2}\right) \sim \frac{1}{2}\left(\frac{\lambda_P}{2an_\infty}\right)^2 \tag{17}$$

This is the most interesting result indicating that the *ratio of two energies depends only on the size of the mode relative to the plasma wavelength.*

Expression (17) is extremely important from the point of view of energy dissipation of the mode. The one and only dissipation mechanism is velocity damping with the rate $\gamma$, which means that the kinetic energy gets damped with the rate $2\gamma$. Since on average, $U_Q(t) = U_K(t) + U_M(t)$ is? one half of total energy, and magnetic energy is obviously not dissipating, the overall effective energy damping rate is [29]

$$\gamma_{eff} = \gamma \frac{U_K}{U_M + U_K} \sim \left(\frac{\lambda_P}{2an_\infty}\right)^2 \Bigg/ \left[\left(\frac{\lambda_P}{2an_\infty}\right)^2 + 2\right] \tag{18}$$

This expression shows *the major difference between the plasmonic and phononic sub-wavelength structures.* For metals the plasma wavelength is about 140-150nm, hence for metallic structures operating in the mid-IR the characteristic dimension may be substantially smaller than the operating (resonant) wavelength $\lambda_0$ but still an order of magnitude larger than $\lambda_P/2an_\infty$ and the effective loss can be orders of magnitude smaller than the metal damping rate $\gamma \sim 10^{14} s^{-1}$. Strictly speaking, when one operates far from plasma frequency ($\lambda \gg \lambda_P$) and the kinetic energy of electrons is small, it is preferable not to invoke the term "plasmon" and instead use the terms "metal waveguides" or "metal nanoantennas". For phononic structures on the other hand the operational wavelength is always much shorter than the plasma wavelength, meaning that in the sub-wavelength phononic structure the effective loss is always equal to the phonon damping rate (which is, as mentioned above, two orders of magnitude less than the electrons damping rate in the metal). Therefore, despite the lower momentum damping rate, it is far from obvious that SPhPs offer any significant advantage over the metallic structures.

3. **Energy balance for self-sustained oscillation**

Now, the self-sustaining eigenmodes can exist only when the in-phase and quadrature energies are equal to each other, i.e. $\int \langle U_I(t) \rangle d^3\mathbf{r} = \int \langle U_Q(t) \rangle d^3\mathbf{r}$ or

$$\frac{1}{4}\int \varepsilon_r' E^2(\mathbf{r})d^3r + \frac{1}{8}\omega\int \frac{\partial \varepsilon_r'}{\partial \omega} E^2(\mathbf{r})d^3r = \frac{1}{8}\omega\int \frac{\partial \varepsilon_r'}{\partial \omega} E^2(\mathbf{r})d^3r + \left(\frac{2na}{\lambda}\right)^2 \frac{1}{4}\int \varepsilon_r' E^2(\mathbf{r})d^3r \qquad (19)$$

Note that we have kept two equal dispersive terms on both sides of (19) in order to obtain solutions from energy balance considerations. Altogether, the expression (19) allows for *four* different classes of solutions, each having different physical origin and characteristics.

To start, there exist *two* possible solutions of (19) that do not depend on the dimensions of the mode. They obviously occur when the dielectric constant is zero, or when dispersive terms on both sides approach the infinity. The *first solution* occurs at $\omega_1 = \omega_{LO}$ and is nothing but the (bulk) *longitudinal optical phonon*. The *second solution* occurs at $\omega_2 = \omega_{TO}$ and is obviously a *transverse optical phonon* mode. Neither one of these solutions is interesting from a practical point of view. In the TO mode the electric field is zero and the electric field inside the LO phonon mode is all contained deep inside the material. The *third possible solution*, occurring when $a \sim \lambda/2n$, is obviously a standard *Fabry-Perot* type mode inside the dielectric resonator (as shown in the Fig.3a).

However, as long as the spatial extent of the mode is sub-wavelength, which in the context of this work means $a < \lambda/2n$, the only way the solution is attainable is when the medium is spatially inhomogeneous and *incorporates regions with both negative and positive dielectric constants* such that

$$\int \varepsilon_r'(\mathbf{r}) E^2(\mathbf{r})d^3r = 0 \;. \qquad (20)$$

The easiest way in which the condition (20), for this *fourth solution*, can be satisfied involves a sub-wavelength object made from a dielectric in the Reststrahlen region $\varepsilon_r'(\omega_0) < 0$ surrounded by cladding made from a "normal" or low-dispersion dielectric with $\varepsilon_{r,cl}'(\omega_0) > 0$, as shown in Fig.3b. Whenever the energy balance condition (19) is satisfied at frequency $\omega_0$, this frequency is the eigen-frequency of the SPhP mode. Obviously, condition (20) can also be satisfied in the sub-wavelength SPP structure shown in Fig.3c which is made from a doped semiconductor such as InAs.

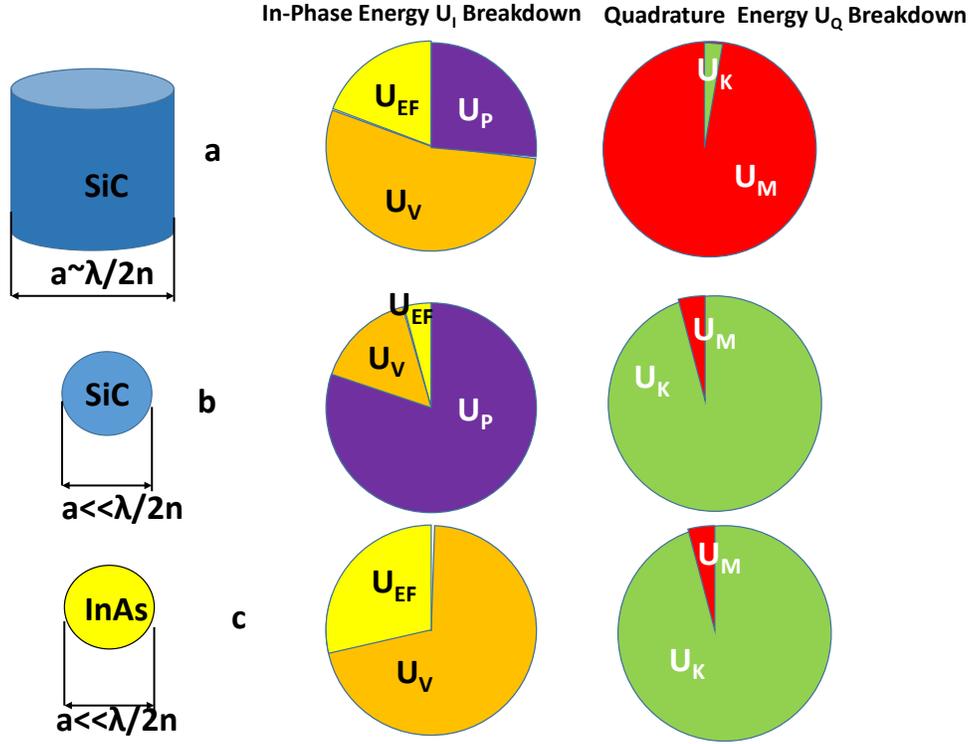

Figure 3 The breakdown of energies in three different resonant structures in the mid-IR region (a) dielectric resonator (b) localized surface phonon polariton (c) localized surface plasmon polariton

Next to each structure we show the breakdown of the in-phase $U_I$ and quadrature $U_Q$ energies. For a typical dielectric resonator operated outside the Reststrahlen (Fig 3a) region the in-phase energy is split between the energies of electric field $U_{EF}$, valence electrons $U_V$, and potential energy of ions $U_P$. For the sub-wavelength SPhP (Fig.2b), the potential energy of ions dominates, but in the SPP shown in Fig.3c this potential energy is absent since free carriers by definition have no potential energy. This dominance vs absence of potential energy of ions constitutes the major difference between SPhPs and SPPs with the repercussions explained in the rest of this work. As far as the quadrature component of energy goes, SPhP and SPP behave similarly, with kinetic energy of ions or electrons $U_K$ dominating the magnetic energy $U_M$ in stark constant to the dielectric resonator energy breakdown of Fig.3a. shown above.

Note that the *in-phase energy breakdown in essence determines the strength of the electric field enhancement* (it increases with the $(U_{EF}/U_I)^{1/2}$ ) and the *quadrature energy breakdown determines the non-radiative loss* (it increases with the $U_K/U_Q$ ) . *The fact that these important conclusions can be inferred from the physically-transparent picture in Fig.3 without resorting to numerical modeling in itself may be considered to be an important result of this work.*

## 4. Localized Surface Phonon Polaritons –enhancing energy and electric field

### 4.a Resonant frequency and Effective volume

To illustrate how the energy breakdown affects various phononic and plasmonic structures consider the most simple example of a spherical particle, with radius $a$ and dielectric constant $\varepsilon_r(\omega)$, placed in the cladding (dielectric constant $\varepsilon_{r,cl}$) as shown in Fig.1c with a dipole moment $\boldsymbol{p} = p\hat{z}$ that is related to the field inside it as

$$\boldsymbol{p} = \frac{4}{3}\pi a^3 \varepsilon_0 \left(\varepsilon_r - \varepsilon_{r,cl}\right) \boldsymbol{E}_{in} \tag{21}$$

The field outside the particle is

$$\boldsymbol{E}_{out}(\boldsymbol{r}) = \frac{1}{4\pi\varepsilon_0 \varepsilon_{r,cl} r^3}\left[3(\boldsymbol{p}\cdot\hat{r})\hat{r} - \boldsymbol{p}\right] = |E_{in}|\frac{a^3}{r^3}\frac{\varepsilon_r - \varepsilon_{r,cl}}{3\varepsilon_{r,cl}}\left[3\cos\theta(\hat{z}\cos\theta + \hat{r}_\perp\sin\theta) - \hat{z}\right] \tag{22}$$

where $\hat{r}_\perp$ is the unit vector in the $xy$ plane. Then one can evaluate the energy integrals outside

$$I_{out} = U_{P1,out} + U_{EF,out} + U_{V,out} = \varepsilon_{r,cl}\int |E_{out}(r,\theta)|^2 d^3\boldsymbol{r} = \frac{8}{3}\pi a^3 \varepsilon_{r,cl}\left[\frac{\varepsilon_r - \varepsilon_{r,cl}}{3\varepsilon_{r,cl}}\right]^2 |E_{in}|^2 \tag{23}$$

and inside

$$I_{in} = U_{P1,in} + U_{EF,in} + U_{V,in} = \int \varepsilon_r |E_{in}(r,\theta)|^2 d^3\boldsymbol{r} = \frac{4}{3}\pi a^3 \varepsilon_r |E_{in}|^2 \tag{24}$$

the particle which according to (20) immediately leads to the eigen-mode condition

$$I_{in} + I_{out} = \varepsilon_r + 2\varepsilon_{r,cl}\left[\frac{\varepsilon_r - \varepsilon_{r,cl}}{3\varepsilon_{r,cl}}\right]^2 = 0 \ . \tag{25}$$

Note that since cladding is normally non-dispersive, $I_{out} > 0$ is the total potential energy in the cladding, but $I_{in} < 0$ is not the complete energy inside the nanoparticle.

Obviously the solution for the dipole mode $\varepsilon_r(\omega) = -2\varepsilon_{r,cl}$ satisfies (25). Thus the solution for SPhP resonant frequency

$$\omega_0^2 = \omega_{TO}^2 + \omega_P^2 /(1 + 2\varepsilon_{r,cl}/\varepsilon_\infty) \tag{26}$$

can be obtained purely from energy conservation considerations. Interestingly enough, the second solution of (25) is $2\varepsilon_r(\omega) = -\varepsilon_{cl}$ and it corresponds to the eigenmode of the spherical void inside the $\varepsilon < 0$ material filled with a conventional $\varepsilon > 0$ dielectric.

One can therefore make an important statement regarding the energy balance in the deeply subwavelength SPhP (and also, of course, SPP) mode. Equation (19) can be re-written as

$$\int \langle U_{P2} \rangle d^3r = \int \langle U_K \rangle d^3r = \frac{1}{8}\omega \int \frac{\partial \varepsilon_r'}{\partial \omega} E_{in}^2(r) d^3r \tag{27}$$

indicating that the energy in the mode oscillates between the kinetic energy inside the dispersive medium (when the phase is 90 degrees or 270 degrees) and the combination of the potential energies of ions and electrons both inside and outside the dispersive medium (when the phase is 0 of 180 degrees). The power constantly flows into the cladding and back. The total energy can then be estimated simply as

$$\int \langle U_E \rangle d^3r = 2\int \langle U_K \rangle d^3r = \frac{1}{4}\omega \int \frac{\partial \varepsilon_r'}{\partial \omega} E_{in}^2(r) d^3r = \frac{2}{3}\pi a^3 \varepsilon_0 \varepsilon_\infty \frac{\omega^2 \omega_P^2}{\left(\omega_{TO}^2 - \omega^2\right)^2 + \omega^2 \gamma^2} E_{in}^2 \tag{28}$$

Substituting the eigen frequency from (26) and using the fact that $\omega_P \gg \gamma$ we obtain

$$\int \langle U_E \rangle d^3r = \frac{2}{3}\pi a^3 \varepsilon_0 \left(\varepsilon_\infty + 2\varepsilon_{r,cl}\right)\left[1 + \frac{\omega_{TO}^2}{\omega_P^2}\left(1 + 2\varepsilon_{r,cl}/\varepsilon_\infty\right)\right] E_{in}^2 = 2\pi a^3 \varepsilon_0 \varepsilon_{r,cl} K_v K_{ph} E_{in}^2, \tag{29}$$

where the $K_v = (\varepsilon_\infty/\varepsilon_{r,cl} + 2)/3$ is the factor describing the "excess" potential energy stored in the valence electrons and

$$K_{ph} = 1 + \frac{\omega_{TO}^2}{\omega_P^2}\left(1 + 2\varepsilon_{r,cl}/\varepsilon_\infty\right) = \frac{\varepsilon_{st} + 2\varepsilon_{r,cl}}{\varepsilon_{st} - \varepsilon_\infty} \tag{30}$$

is the factor corresponding to the energy stored in, and oscillations of the ions, while $\varepsilon_{st}$ is the static dielectric constant. Using (26) one can also express the "excess potential energy factor" as

$$K_{ph} = \left(\omega_0^2/\omega_P^2\right)\left(1 + 2\varepsilon_{r,cl}/\varepsilon_\infty\right) \tag{31}$$

Obviously, for the SPP this factor is equal to unity, but for a typical phononic medium, such as SiC with $\varepsilon_{st} = 9.7$, $\varepsilon_\infty = 6.7$, and dielectric cladding with $\varepsilon_{r,cl} = 1$ one obtains $K_{ph} \sim 4$.

Next one can introduce the effective volume as

$$V_{eff} = \frac{\int \langle U_E \rangle d^3r}{\frac{1}{4}\varepsilon_0 \varepsilon_{r,cl} |E_{max}|^2} = \frac{3}{2} V_n K_v K_{ph} \tag{32}$$

where the maximum field is $E_{max} = 2E_{in}$ and $V_n = 4/3\pi a^3$ is the nanoparticle volume. The first term, $K_v$ indicates that a significant part of the energy is stored in the oscillation of valence electrons, and the second term $K_{ph}$ indicates that an even larger fraction of the energy is stored in the lattice vibration. As a result, even when the *energy* can be efficiently concentrated by the nanoparticle, the enhancement of the *electric field* may be less than stellar.

### 4.b Electric field Enhancement

We now turn our attention to the estimates of the field enhancement. In the presence of the external electric field $E_{ext}$ the nanoparticle acquires the dipole moment

$$p = 4\pi\varepsilon_0 \varepsilon_{r,cl} a^3 E_{ext} \frac{\varepsilon_r - \varepsilon_{r,cl}}{\varepsilon_r + 2\varepsilon_{r,cl}} \tag{33}$$

Substituting (1) we obtain in the vicinity of resonance $\omega \approx \omega_0$ (26)

$$p \approx 4\pi\varepsilon_0 \varepsilon_{r,cl}^2 \varepsilon_\infty a^3 \frac{3}{(\varepsilon_\infty + 2\varepsilon_{cl})^2} \frac{\omega_P^2}{\omega_0^2 - \omega^2 - j\omega\gamma} E_{ext}$$

According to (22)

$$E_{max} = \frac{p}{2\pi\varepsilon_0 \varepsilon_{r,cl} a^3} = \varepsilon_{r,cl}\varepsilon_\infty \frac{6}{(\varepsilon_\infty + 2\varepsilon_{cl})^2} \frac{\omega_P^2}{\omega_0^2 - \omega^2 - j\omega\gamma} E_{ext} \tag{34}$$

Now, if we introduce the quality factor $Q_0 = \omega_0 / \gamma$, then at resonance we obtain

$$E_{max} = \varepsilon_{r,cl}\varepsilon_\infty \frac{6}{(\varepsilon_\infty + 2\varepsilon_{cl})^2} \frac{\omega_P^2}{\omega_0^2} Q_0 E_{ext} = \frac{2Q_0}{K_v K_{ph}} E_{ext} = \frac{E_{max,0}}{K_v K_{ph}} \tag{35}$$

where $E_{max,0} = 2Q_0 E_{ext}$ is the enhancement achieved near the nanosphere made from a "classical Drude" metal with $\varepsilon_\infty = 1$ placed into the vacuum $\varepsilon_{r,cl} = 1$. As one can see, the same two factors are responsible for the decrease in the enhancement of the field compared to the Drude metal with $\varepsilon_\infty = 1$. Now we can use (32) and (35) to find the energy inside the SPhP mode

$$\int \langle U_E \rangle d^3 r = \frac{1}{4}\varepsilon_0 \varepsilon_{r,cl} E_{max}^2 V_{eff} = 3V_n \frac{Q_0^2}{K_v K_{ph}} \langle U_{ext} \rangle \tag{36}m$$

where the external energy density is $\langle U_{ext} \rangle = \frac{1}{2}\varepsilon_0 \varepsilon_{r,cl} E_{ext}^2$. The energy in the mode is reduced by the factor $K_v K_{ph}$ because the effective dipole is smaller in SPhP when compared to the Drude metal. On top of that, the effective volume is larger in SPhP by the same factor – hence the *energy density* gets enhanced by a factor $(K_v K_{ph})^2$ less compared to the Drude metal with $\varepsilon_\infty = 1$.

To illustrate these results we compare performance of two spherical structures, the first one is a SiC sphere with a 0.5μm diameter and the other is the sphere of equal size comprised of $In_{0.53}Ga_{0.47}As$ n-doped with the donor density of $N_d=6.5 \times 10^{18} cm^{-3}$. The scattering rate in SiC, as previously mentioned, is $\gamma \approx 1.1 ps^{-1}$ while for InGaAs it has been estimated from the mobility data [45,46] to be about 8 times higher, $\gamma_s \approx 8.5 \times ps^{-1}$. The results are shown in Fig. 4 – the resonance of the phononic structure has a very narrow FWHM linewidth of about 0.28 THz (Q~100) compared to InGaAs – 2.2 THz (Q~13) yet the field enhancement achieved with phononic structure (about 34-fold) is only three times higher than in the plasmonic structure (about 11-fold). This discrepancy is due to the fact that most of energy is stored in the lattice vibrations in SiC. The difference in the enhancement would have been even less than a factor of 3 if not for the fact that the larger fraction of energy is stored in the oscillations of valence electrons in InGaAs ($K_v = 4.5$ for InGaAs vs. $K_v = 2.9$ for SiC).

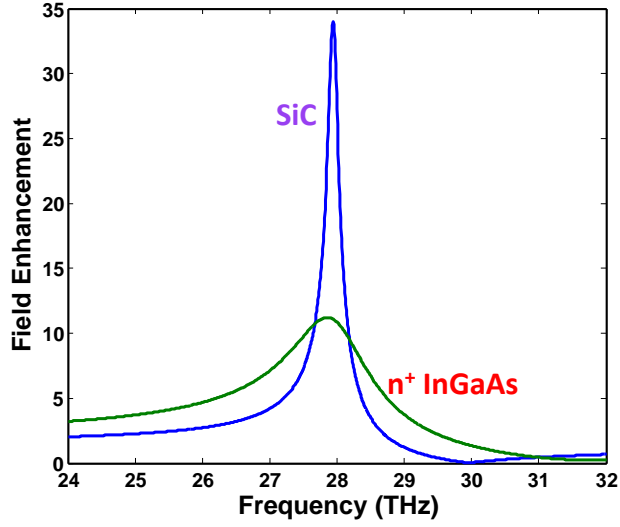

Figure 4. Maximum field enhancement provided by the 0.5 µm spheres of SiC and n-doped InGaAs

### 4.c Purcell enhancement

It is also interesting to see by how much one can enhance the spontaneous emission rate [47-49] $\gamma_0$. The Purcell factor that describes enhancement relative to the emission into an infinite dielectric can be found as [50,51]

$$F_P = \frac{3}{4\pi}\left(\frac{\lambda}{n_{cl}}\right)^3 V_{eff}^{-1} \frac{\omega}{\gamma_R + \gamma} = \frac{2}{3\pi}\left(\frac{\lambda}{n_{cl}}\right)^3 \frac{V_n^{-1}}{K_{env}K_{ph}} \frac{\omega}{\gamma_R + \gamma} \qquad (37)$$

The radiative decay of the dipole (21) in turn can be found as [48]

$$\gamma_R = 4\pi^2 \frac{V_n}{K_{env}K_{ph}}\left(\frac{n_{cl}}{\lambda}\right)^3 \omega \qquad (38)$$

If we in factor

$$F_{P,eff} = \frac{8\pi}{3}\frac{1}{\left(K_{env}K_{ph}\right)^2} \frac{1}{\left(4\pi^2 \frac{V_n'}{K_{env}K_{ph}}\right)^2 + \frac{1}{Q_0^2}} \leq \frac{8\pi}{3}\frac{Q_0^2}{\left(K_{env}K_{ph}\right)^2}. \qquad (39)$$

This is precisely the same result as shown for the field enhancement. One should note that typically the Q-factor for SiC is on the scale of 200 or so vs. 20-30 for InAs. Therefore, both field enhancement and Purcell factor characteristics in SPhPs and SPPs operating in the mid IR region differ only by a factor of a few, despite nearly an order of magnitude longer scattering times for phonons.

## 5. Propagating Surface Phonon Polaritons

Let us now consider a propagating polariton between two media: one with real positive dielectric constant with weak dispersion $\varepsilon_{r,cl}$ (Fig.1b) and one with the dispersive complex dielectric constant $\varepsilon_r(\omega)$ corresponding to either phononic (1) or plasmonic (2) material. The dispersion of the propagation constant of the polariton at the interface is

$$\beta^2(\omega) = k_{cl}^2(\omega) \frac{\varepsilon_r(\omega)}{\varepsilon_r(\omega) + \varepsilon_{r,cl}} \tag{40}$$

where $k_{cl}^2(\omega) = \varepsilon_{r,cl}^{1/2}\omega/c$. Differentiating (40) over frequency and assuming that one operates near the resonance, i.e. $\varepsilon_r'(\omega) \approx -\varepsilon_{r,cl}$ (see Supplementary Material for the details), one can obtain the expression for group velocity

$$v_g^{-1} = \frac{d\beta}{d\omega} = \frac{1}{v_{cl}} \frac{\beta}{k_{cl}} \left[ 1 + \frac{\beta^2}{k_{cl}^2} \frac{\varepsilon_{r,cl}}{\varepsilon_\infty} \frac{\omega^2 \omega_P^2}{(\omega_{LO}^2 - \omega^2)^2} \right] \approx \frac{1}{v_{cl}} \frac{\beta}{k_{cl}} \left[ 1 + \frac{3}{4} \frac{\beta^2}{k_{cl}^2} K_v K_{ph} \right] \tag{41}$$

where $v_{cl}$ is the propagation velocity in the cladding, and the factors $K_v = (2 + \varepsilon_\infty/\varepsilon_{r,cl})/3$ and $K_{ph} = (1 + 2\varepsilon_{r,cl}/\varepsilon_\infty)\omega_0^2/\omega_P^2$ have been previously defined in (31).

As expected, the group velocity in the SPhP is reduced by an additional factor of $K_{ph}$ when compared to the SPP because most of the energy is not associated with the photon but with the lattice vibrations, which have very low group velocity. Note that the factor $\beta/k_{cl}$ in front of the brackets is the reduction of the phase velocity of the SPhP and is associated with the reduction of the magnitude of the magnetic field and Poynting vector inside the cladding which causes a decrease of the energy propagation velocity. The quadratic factor $\beta^2/k_{cl}^2$ inside the brackets is also associated with the fact that the Poynting vector inside the negative permittivity medium is directed backward.

Also from (41) we can obtain the expression for the imaginary part of propagation function

$$\beta'' \approx \frac{\partial \beta}{\partial \varepsilon} \varepsilon'' = \frac{\partial \beta}{\partial \omega} \left( \frac{\partial \varepsilon}{\partial \omega} \right)^{-1} \varepsilon'' = \frac{1}{2} \gamma v_g^{-1} \tag{42}$$

This means that the propagation length of the SPhP is

$$L_p = 1/2\beta'' = v_g/\gamma = \frac{v_{cl}\gamma^{-1}}{\frac{\beta}{k_{cl}}\left[1 + \frac{3}{4}\frac{\beta^2}{k_{cl}^2}K_v K_{ph}\right]} \approx \frac{\lambda_{cl}}{2\pi} \frac{4Q_0}{3(\beta/k_{cl})^3 K_v K_{ph}} . \tag{43}$$

Once again, the factor $Q_0/K_{ph}$ makes an appearance, indicating that enhancement of the propagation length in SPhPs is small relative to SPPs, despite a much lower scattering rate. The SPhPs definitely live longer than SPPs but since they propagate slower their propagation length is not as long as one would expect.

To illustrate these results we have considered the waveguides Fig.1b with air cladding and the SiC or n-doped ($N_d=6.2 \times 10^{18} cm^{-3}$) InGaAs as a negative $\varepsilon$ material.

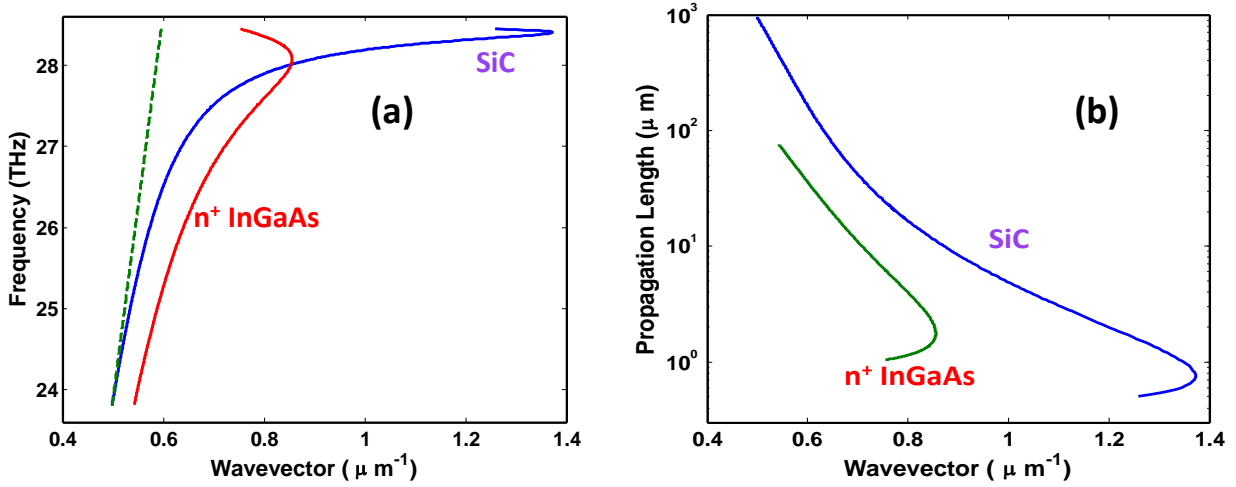

Figure 5 (a) dispersion of propagating phonon (SiC) and plasmon (InGaAs) polaritons. The dashed line represents the light line in the cladding. (b) Propagating lengths vs. wavevector of the same.

In Fig.5a the dispersion curves are shown. Due to its slower scattering rate, the SPhP in SiC reaches larger wavevectors (its effective index $\beta/k_0$ reaches maximum value of 2.8) and has a longer propagation length than SPPs in n-doped InGaAs, as can be seen in Fig.5b. However, the difference between the two is not as high as one could expect from the nearly order of magnitude difference between the scattering rates. This can be explained by the fact that in SiC a significant part of the energy is contained in oscillations of phonons, which slows down the propagation velocity and thus increases the loss.

## 6. Comparison with guided bulk phonon polaritons

Having investigated the surface phonon polaritons and how they compare with surface plasmon polaritons one can go back to our discussions in Section 3 where we have discussed the means of obtaining self-sustaining modes in small volumes, shown in Fig.3. So far we have only considered the structures combining regions with positive and negative dielectric constants, i.e. interface polaritons as represented in Fig.3(b). However, what about operating with the same polar material in the "dielectric regime" (Fig.3a) where the dielectric constant remains positive but large as the frequency approaches $\omega_{TO}$? In this case the photons get coupled with transverse optical phonons forming bulk phonon polaritons which then can be confined in either a 3-dimensional resonant structure (Fig.3a) or in a slab waveguide structure as a guided bulk phonon polaritons polariton (GBPhP) mode (inset of Fig.6a). The 3D structure can support a Fabry-Perot type mode capable of tight confinement but it suffers from high radiative losses at the facets, the wave guiding structure has no such problem.

We therefore consider a slab waveguide structure of Fig.6a made of the polar dielectric with positive dielectric constant $\varepsilon_r(\omega)$ as in (1) surrounded by the non-dispersive cladding dielectric $\varepsilon_{r,cl} < \varepsilon_r(\omega)$, which can be air. One can confine the light fairly well in the waveguide with thickness of about $w \sim \lambda/2\varepsilon_r^{1/2}(\omega) = \pi/\beta(\omega)$ where

$$\beta(\omega) = k_0 \varepsilon_r^{1/2}(\omega) = (\omega/c)\varepsilon_r^{1/2}(\omega) \tag{44}$$

Performing differentiation of (44) we obtain the expression for the group velocity of GBPhP (see Supplementary Material for the details)

$$v_g^{-1} = \frac{1}{c}\frac{\beta}{k_{cl}}\left[1 + \varepsilon_\infty^{-2}\frac{\beta^2}{k_0^2}\frac{\omega^2 \omega_P^2}{(\omega_{LO}^2 - \omega^2)^2}\right] \approx \frac{1}{c}\frac{\beta}{k_0}\left[1 + \varepsilon_\infty^{-2}\frac{\beta^2}{k_0^2}\frac{\omega^2}{\omega_P^2}\right] \tag{45}$$

which immediately gives the value of propagation length for the guided polariton

$$L_p = v_g/\gamma = \frac{c\gamma^{-1}}{\frac{\beta}{k_0}\left[1 + \varepsilon_\infty^{-2}\frac{\beta^2}{k_0^2}\frac{\omega^2}{\omega_P^2}\right]} \approx \frac{\lambda_0}{2\pi}\frac{Q_0 \varepsilon_\infty^2}{(\beta/k_0)^3 \omega^2/\omega_P^2} . \tag{46}$$

Comparison of propagation lengths of the guided (46) and surface (43) polaritons propagating with the same effective index $\beta/k_0$ assuming an air cladding yields

$$L_{p,guided}/L_{p,surface} \simeq 3\varepsilon_\infty^3(\tfrac{1}{4} + \varepsilon_\infty^{-1} + \varepsilon_\infty^{-2}) . \tag{47}$$

Therefore, for similar confinement guided polaritons an improved enhanced propagation length is obtained. This can be related to the fact that in guided polaritons the energy propagates forward both in the guide and in the cladding, while regarding surface polaritons the energy in the material with negative epsilon propagates backward.

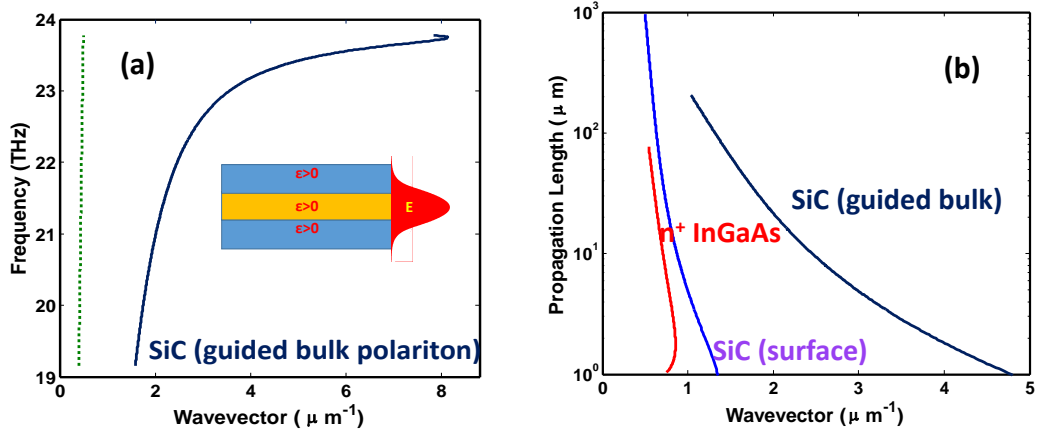

Figure 6 (a) dispersion of guided bulk phonon polariton in SiC shown in the inset Dashed line represents the light line in the cladding. (b) Propagating lengths vs. wavevector of the guided polariton compared with the surface phonon polariton and surface plasmon.

In Fig. 6a the dispersion of the GBPhP mode in SIC is shown – as expected, far larger wavevectors can be attained at frequencies slightly below the TO frequency than what is achievable with SPhPs as shown on Fig.5a. Also, as shown in Fig.6b the propagation length of GBPhP greatly exceeds that of either SPhP or SPP for the same wavevector, i.e. for comparable confinement inside the cladding.

When it comes to propagating polaritons one should also always remember that a simple metal-insulator-metal (MIM) waveguide is capable of supporting relatively low loss propagation at IR wavelengths. According to [29] an Au MIM waveguide with a 1µm gap (confinement similar to the one in SiC polariton waveguide of Fig 6) will have propagation length off about 100 µm for the wavelength in the 10-12 µm range, i.e. at least as good as polariton waveguide. This is probably the main factor that will determine usefulness of phononics: *phononic devices are indeed offering reduced loss in the mid to far IR range, but in this range the all-metal structures (such as MIM guides and patch antennas) are also relatively low loss despite the intrinsically high loss of metal because the field simply does not penetrate the metal beyond roughly 100nm skin depth.*

Therefore, as one considers going to wavelengths longer than 10-12 µm using various III-V and II-VI polar materials, the relative advantages of metal structures become more and more prominent. This is of course a well-known fact. For instance, while quantum cascade lasers operating in the mid-IR range do use dielectric waveguides, the ones operating in far IR (52) and THz (53-55) regions of the spectra always rely upon double clad metal waveguides to achieve maximum confinement at low loss.

## 7 Conclusions

Using a simple yet physically insightful energy balance model (rather than relying on tedious numerical simulation) we have shown that although the scattering rate of optical phonons in polar dielectrics is an order-of-magnitude less than the scattering rate of electrons in doped semiconductors and two orders of magnitude less than in metals, this advantage is counter-balanced by the smaller effective plasma frequency and the fact that a significant part of energy is stored in the potential energy of lattice vibrations. Therefore, phononics, while being a valuable technology does not hold an overwhelming advantage over the metal plasmonics in the long wavelength region of the spectrum. Since phononics is limited to a relatively narrow spectral region either below or above the TO phonon frequency, while plasmonic structures have design flexibility allowing them to operate over any spectral region, the question of whether to use metal structures, semiconductor plasmonics, or phononics should be handled judiciously for each specific application.

## Acknowledgements

This material is based upon work supported by the National Science Foundation under Grant No. DMR 1507749 and Army Research Office Grant W911NF-15-1-0629. As ever, Prof. P. Noir's support and fruitful discussions with him have proven invaluable.

# Relative merits of Phononics vs. Plasmonics: the energy balance approach.

# Supplementary Information

## Derivation of the expression for group velocity of the SPhP (Eq.41)

We start with the expression (40)

$$\beta^2(\omega) = k_{cl}^2(\omega) \frac{\varepsilon_r(\omega)}{\varepsilon_r(\omega) + \varepsilon_{r,cl}} \qquad (1)$$

and re-write it as

$$\frac{\beta^2(\omega)}{k_{cl}^2(\omega)} = \frac{\varepsilon_r(\omega)}{\varepsilon_r(\omega) + \varepsilon_{r,cl}} \qquad (2)$$

Next we differentiate both sides of (2) over frequency

$$\frac{2\beta \frac{d\beta}{d\omega} k_{cl}^2 - 2\beta^2 k_{cl} \frac{dk_{cl}}{d\omega}}{k_{cl}^4} = \frac{\frac{\partial \varepsilon_r'}{\partial \omega}\left(\varepsilon_r(\omega)+\varepsilon_{r,cl}\right) - \varepsilon_r(\omega)\frac{\partial \varepsilon_r'}{\partial \omega}}{\left(\varepsilon_r(\omega)+\varepsilon_{r,cl}\right)^2} = \frac{\frac{\partial \varepsilon_r'}{\partial \omega}\varepsilon_{r,cl}}{\left(\varepsilon_r(\omega)+\varepsilon_{r,cl}\right)^2} \qquad (3)$$

Assuming that group and phase velocities in cladding are the same, i.e. $dk_{cl}/d\omega \approx k_{cl}/\omega = v_{cl}$ we obtain

$$k_{cl}\frac{d\beta}{d\omega} - \beta\frac{k_{cl}}{\omega} = \frac{k_{cl}^3}{2\beta}\frac{\varepsilon_{r,cl}}{\left(\varepsilon_r(\omega)+\varepsilon_{r,cl}\right)^2}\frac{\partial \varepsilon_r'}{\partial \omega} \qquad (4)$$

Next take the square of Eq (2) as

$$\frac{k_{cl}^4}{\left(\varepsilon_r(\omega)+\varepsilon_{r,cl}\right)^2} = \frac{\beta^4}{\varepsilon_r^2(\omega)} \qquad (5)$$

and substitute it it into the r.h.s. of (4) to obtain

$$k_{cl}\frac{d\beta}{d\omega} - \beta\frac{k_{cl}}{\omega} = \frac{1}{2}\frac{\beta^3 \varepsilon_{r,cl}}{k_{cl}\varepsilon_r^2(\omega)}\frac{\partial \varepsilon_r'}{\partial \omega} \qquad (6)$$

and then

$$\frac{d\beta}{d\omega} - \frac{1}{v_{cl}}\frac{\beta}{k_{cl}} = \frac{1}{2}\frac{\beta^3 \varepsilon_{r,cl}}{k_{cl}^2 \varepsilon_r^2(\omega)}\frac{\partial \varepsilon_r'}{\partial \omega} = \frac{1}{2v_{cl}}\frac{\beta^3 \varepsilon_{r,cl}}{k_{cl}^3 \varepsilon_r^2(\omega)}\frac{\omega \partial \varepsilon_r'}{\partial \omega} \qquad (7)$$

Which leads to the expression for the inverse group velocity

$$v_g^{-1} = \frac{d\beta}{d\omega} = \frac{1}{v_{cl}}\frac{\beta}{k_{cl}}\left[1 + \frac{1}{2}\frac{\beta^2 \varepsilon_{r,cl}}{k_{cl}^2 \varepsilon_r^2(\omega)}\frac{\omega \partial \varepsilon_r'}{\partial \omega}\right] \qquad (8)$$

Once we calculate the dispersive term

$$\frac{\omega \partial \varepsilon_r^{'}}{2\varepsilon_r^2 \partial \omega} = \frac{\omega^2 \omega_P^2}{\varepsilon_\infty \left(\omega_{TO}^2 - \omega^2\right)^2} \bigg/ \left(1 + \frac{\omega_P^2}{\left(\omega_{TO}^2 - \omega^2\right)}\right)^2 = \frac{\omega^2 \omega_P^2}{\varepsilon_\infty \left(\omega_{LO}^2 - \omega^2\right)^2} \qquad (9)$$

We indeed obtain

$$v_g^{-1} = \frac{1}{v_{cl}} \frac{\beta}{k_{cl}} \left[1 + \frac{\beta^2}{k_{cl}^2} \frac{\varepsilon_{r,cl} \omega^2 \omega_P^2}{\varepsilon_\infty \left(\omega_{LO}^2 - \omega^2\right)^2}\right] \qquad (10)$$

which is equivalent to the first equality of Eq (40) in the main text.

To proceed further we note that $\omega \approx \omega_0$ i.e.

$$\omega^2 = \frac{\omega_{LO}^2 + \left(2\varepsilon_{r,cl}/\varepsilon_\infty\right)\omega_{TO}^2}{1 + \left(2\varepsilon_{r,cl}/\varepsilon_\infty\right)} \qquad (11)$$

And also

$$\omega_{LO}^2 - \omega^2 = \frac{\omega_P^2 \, 2\varepsilon_{r,cl}}{2\varepsilon_{r,cl} + \varepsilon_\infty} \qquad (12)$$

Well then

$$\frac{\varepsilon_{r,cl} \omega^2 \omega_P^2}{\varepsilon_\infty \left(\omega_{LO}^2 - \omega^2\right)^2} \approx \frac{\varepsilon_{r,cl} \omega^2}{\varepsilon_\infty \omega_P^2} \cdot \frac{\left(2\varepsilon_{r,cl} + \varepsilon_\infty\right)^2}{4\varepsilon_{r,cl}^2} = \frac{\omega^2}{\omega_P^2} \frac{\left(2\varepsilon_{r,cl} + \varepsilon_\infty\right)^2}{4\varepsilon_\infty \varepsilon_{r,cl}} = \frac{\omega^2}{\omega_P^2} \frac{\left(1 + 2\varepsilon_{r,cl}/\varepsilon_\infty\right)\left(2 + \varepsilon_\infty/\varepsilon_{r,cl}\right)}{4} = \frac{3}{4} K_v K_{ph} \qquad (13)$$

Which confirms the second equality in (40).

**Derivation of the expression for the group velocity of the guided bulk polariton mode (Eq.45)**

We start with Eq (44) in the main text written as

$$\frac{\beta^2(\omega)}{k_0^2(\omega)} = \varepsilon_r(\omega) \qquad (14)$$

and differentiate it over frequency. Derivative of the left side yields

$$\frac{d}{d\omega}\left(\frac{\beta^2}{k_0^2}\right) = \frac{2\beta \frac{d\beta}{d\omega} k_0^2 - 2\beta^2 k_0 \frac{dk_0}{d\omega}}{k_0^4} = \frac{2\beta}{k_0^2}\left(\frac{d\beta}{d\omega} - \frac{\beta}{\omega}\right) = \frac{2\beta}{k_0^2}\left(\frac{d\beta}{d\omega} - \frac{\beta}{k_0} \frac{1}{c}\right) \qquad (15)$$

and derivative of the r.h.s. yields

$$\frac{\partial \varepsilon_r'}{\partial \omega} = \frac{2\omega \omega_P^2}{(\omega_{TO}^2 - \omega^2)^2} \times \frac{(\omega_{LO}^2 - \omega^2)^2}{(\omega_{LO}^2 - \omega^2)^2} = \frac{2\omega^2 \omega_P^2}{\omega(\omega_{LO}^2 - \omega^2)^2} \frac{\varepsilon_r^2}{\varepsilon_\infty^2} =$$
$$= \frac{2\omega^2 \omega_P^2}{\omega \varepsilon_\infty^2 (\omega_{LO}^2 - \omega^2)^2} \frac{\beta^4}{k_0^4} = \frac{2\beta}{k_0^2} \cdot \frac{\beta}{k_0 c} \cdot \frac{\beta^2}{k_0^2} \cdot \frac{\omega^2 \omega_P^2}{\varepsilon_\infty^2 (\omega_{LO}^2 - \omega^2)^2}$$

(16)

Then we get

$$\frac{d\beta}{d\omega} = \frac{1}{c} \frac{\beta}{k_0} \left(1 + \frac{\beta^2}{\varepsilon_\infty^2 k_0^2} \frac{\omega^2 \omega_P^2}{\varepsilon_\infty^2 (\omega_{LO}^2 - \omega^2)^2}\right) \approx \frac{1}{c} \frac{\beta}{k_0} \left(1 + \frac{\beta^2}{\varepsilon_\infty^2 k_0^2} \frac{\omega^2}{\omega_P^2}\right)$$

(17)

Where in the last step we assumed that $\omega \sim \omega_{TO}$ and thus obtained Eq.45 in the main text